\documentclass[12pt]{article}
\usepackage{amssymb,amsmath}
\usepackage{comment}
\usepackage{graphicx}
\usepackage{bm}
\usepackage{color}
\usepackage{here}
\usepackage{enumerate}
\usepackage{subfigure}
\numberwithin{equation}{section}

\begin{document}

\thispagestyle{empty}
\begin{flushright}
NCTS-TH/1708
\\

\end{flushright}
\vskip2.5cm
\begin{center}
{\Large 
\bf 
Implications of Conformal Symmetry \\ 
in Quantum Mechanics}

\vskip2cm
Tadashi Okazaki\footnote{tadashiokazaki@phys.ntu.edu.tw} 

\bigskip
{\it 
Department of Physics and Center for Theoretical Sciences,\\
National Taiwan University, Taipei 10617, Taiwan}
\\

\end{center}

\vskip2cm
\begin{abstract}
In conformal quantum mechanics with the vacuum of a real scaling dimension 
and with a complete orthonormal set of energy eigenstates 
which is preferable under the unitary evolution, 
the dilatation expectation value between energy eigenstates 
monotonically decreases along the flow from the UV to the IR. 
In such conformal quantum mechanics there exist bounds on scaling dimensions 
of the physical states and the gauge operators. 
\end{abstract}

\newpage
\setcounter{tocdepth}{2}
\tableofcontents

\section{Introduction}
A space-time symmetry is of great importance 
in defining theories to describe the laws of nature. 
By reconciling quantum mechanics with the Poincar\'{e} symmetry group, 
one obtains relativistic quantum field theory in which 
fields are classified by the Casimir operators of the Poincar\'{e} group 
as quantum numbers, i.e. the spin and the mass.   
In conformal field theory, 
operators are classified 
by the Casimir operators of the conformal group. 
Since the mass cannot be invariant under the conformal group 
as it is a dimensional parameter, it is replaced by the scaling dimension. 
The main subject in the study of conformal field theory 
is the operators which have a definite spin and scaling dimension. 
By virtue of symmetry properties of the correlation functions, 
even if we have no Lagrangian description, 
we can study conformal field theory 
from consistency conditions, via the conformal bootstrap method 
\cite{Ferrara:1973yt,Polyakov:1974gs,Mack:1975jr}, 
which leads to a certain spectrum of the allowed scaling dimensions 
of the theory. 

In this work we investigate conformal quantum mechanics,   
which is obtained 
by the incorporation of the $SL(2,\mathbb{R})$ 
conformal symmetry of time into quantum mechanics. 
In conformal quantum mechanics, 
the Casimir operators of the conformal group $SL(2,\mathbb{R})$ 
is just associated with the scaling dimensions. 
The important fact is that 
while in the analysis of the spectrum for conformal field theory 
the dilatation is identified with the Hamiltonian in the radial quantization, 
in conformal quantum mechanics they should be distinguished. 
Therefore conformal quantum mechanics would be a theoretically appropriate playground to study 
the relationship between the energy and the scaling dimension. 

Here we aim to address two main issues in conformal quantum mechanics 
that admits the vacuum and the primary operators by examining its correlation functions.

The first question is the irreversibility of the RG flow 
in quantum mechanical system. 
The RG transformation is associated 
to the scaling process from the UV to the IR. 
The information in the UV of the theory is expected to be lost in the IR 
since correlation functions in field theory 
are valid only at scales smaller than the cutoff by integrating out higher momentum degrees of freedom. 
Zamolodchikov's $c$-theorem \cite{Zamolodchikov:1986gt} 
proves the irreversibility of the RG flow in two dimensional field
theories by establishing the existence of a $c$-function, 
a function of coupling constant and energy scale 
that decreases along the RG flow 
and is stationary at the RG fixed points, 
where its value equals the crucial parameter 
in the theory, i.e. central charge. 
For even $d$-dimensional conformal field theory (CFT$_{d}$), 
Cardy \cite{Cardy:1988cwa} proposes a $c$-function 
as the quantity appearing in the trace anomaly 
given by the one point function 
of the trace of the energy momentum tensor on the sphere; 
$c\sim \langle T^{\mu}_{\mu}\rangle_{S^{d}}$. 
There have been various proposals 
in higher dimensional extensions of a $c$-function; 
$d=3$ \cite{Myers:2010tj,Jafferis:2010un,Casini:2011kv}, 
$d=4$ \cite{Komargodski:2011vj} 
and $d=6$ \cite{Elvang:2012st, Grinstein:2014xba}. 
An exploration of a $c$-function in conformal quantum mechanics 
is attractive since it may provide a direct explanation 
of the irreversibility 
as the theory distinguishes the energy and the scaling dimension. 
We observe that the trace part of the energy momentum tensor 
is related to the conserved current of the dilatation $D$ 
and we show that the expectation value 
of the dilatation between preferable energy eigenstates under the unitary evolution, 
which we will call a $D$-function $D(E)\sim \langle E|D|E\rangle$ 
possesses similar properties as a $c$-function.

In the second part, 
we present a pair of simple no-go theorems. 
In relativistic quantum field theory, 
interaction of massless particles with spins is highly constrained due to the Lorentz symmetry. 
For higher spin massless particles 
the negative energy state and negative probability occur. 
Even though they are mathematically well-defined, 
they would not appear in experimental results. 
The Weinberg-Witten theorem \cite{Weinberg:1980kq} 
claims that 
a theory with the Lorentzian covariant conserved current $J^{\mu}$ does not 
admit the spin $s>\frac12$ massless charged particles  
and that a theory with the Lorentzian covariant energy momentum tensor $T^{\mu\nu}$ 
does not involve spin $s>1$ massless particles. 
The theorem is proven by investigating the matrix elements 
of the charge operators between one massless particle states 
of spin and momenta. 
In this work we apply a similar method to matrix elements 
in conformal quantum mechanics. 
We derive new types of the no-go theorems 
which impose constraints on scaling dimensions 
of the physical states constructed in terms of the primary operators 
and the preferred vacua under the unitary evolution as 
$-\frac12 \le d+\Delta\le\frac12$. 
This admits the free bosonic scalar with $d+\Delta= -\frac12$, 
the free fermion with $d+\Delta=0$ 
and the bosonic auxiliary field with $d+\Delta=\frac12$. 
We also find a bound on scaling dimension of the gauge operators which may couple to 
physical states with the scaling dimensions $d+\Delta$ as 
$0<\delta_{Q}\le 2$. 
While bosonic scalar of $d+\Delta=-\frac12$ may couple to the gauge fields of $\delta_{Q}=2$, 
the fermion of $d+\Delta=0$ can interact with those of $\delta_{Q}=1$.

The organization of this paper is as follows. 
In section \ref{confsec} 
we review the $SL(2,\mathbb{R})$ conformal symmetry of time. 
In section \ref{vsubsec} 
we discuss the vacuum state $|\Omega\rangle$ 
which can be labeled by the scaling dimension $d$. 
It characterizes the theory under consideration. 
In section \ref{dsec} 
we study the energy eigenstate $|E\rangle$ 
and introduce the $D$-function 
as the dilatation expectation value between the energy eigenstates. 
We see that the $D$-function behaves as a $c$-function. 
In section \ref{nogosec} 
we derive the no-go theorems. 
They provide bounds on scaling dimensions 
of the vacuum, of the primary operator 
and of the gauge operator as well as constraints. 
In section \ref{dissec} we conclude with some open questions and future directions.

\section{Conformal Symmetry of Time}
\label{confsec}
In general a conformal group is the 
group of space-time transformations that preserve angles locally between 
two distinct points, whose definition contains a metric tensor. 
Such definition does not seem to be suitable for one-dimension 
since there are neither angle nor metric tensor. 
In addition, if we follow the definition, 
we find a diffeomorphism group $\mathrm{Diff}(\mathbb{R})$, 
which requires a gravity coupling. 
It does not seem realistic to consider the quantum mechanical 
system which is invariant under the $\mathrm{Diff}(\mathbb{R})$, 
that is the topological quantum mechanics. 
In this work we are interested in the conformal symmetry transformation 
of time which consists of the translation, the scaling transformation 
and the conformal boost. 
The corresponding generators are 
the Hamiltonian $H=i\partial_{t}$, 
the dilatation $D=it\partial_{t}$ 
and the special conformal transformation $K=it^{2}\partial_{t}$ 
respectively. 
Let 
\begin{align}
\label{c1a1}
G&=uH+vD+wK
\end{align}
be a linear combination of the three generators 
where $u$, $v$, $w \in \mathbb{C}$ are constant parameters. 
It then turns out that 
\begin{align}
\label{c1a2}
G&=i\frac{d}{d\tau}, \\
\label{c1a3}
d\tau&=\frac{dt}{u+vt+wt^{2}}. 
\end{align}
Hence the $G$ can be viewed as the new Hamiltonian of the new time coordinate $\tau$. 
From (\ref{c1a3}) the new time coordinate $\tau$ can be represented as 
\begin{align}
\label{c1a4}
\tau&=\int d\tau
=\int_{t_{0}}^{t}\frac{dt'}
{u+vt'+wt^{'2}}+\tau_{0}
\end{align}
where $\tau_{0}=\tau(t_{0})$. 
For simplicity we set $\tau_{0}=0$. 
For example, 
the new time coordinates $\tau$ generated by 
$H$, $D$ and $K$ are as follows:
\begin{enumerate}

\item time evolution

An evolution of the original coordinate $t$ 
is generated by the Hamiltonian $H$.  
The corresponding new time coordinate (\ref{c1a4}) is
\begin{align}
\label{c1b1a}
\tau&=\int^{t}_{t_{0}}
dt'=t-t_{0}.
\end{align}
The finite transformation of the original time coordinate is 
\begin{align}
\label{c1b1b}
t&=t_{0}+\tau
\end{align}
and the infinitesimal transformation is 
\begin{align}
\label{c1b1c}
\delta t&=t-t_{0}=\tau. 
\end{align}

\item global time dilation

The new time coordinate (\ref{c1a4}) 
whose evolution is generated by the dilatation $D$ is 
\begin{align}
\label{c1b2a}
\tau&=\int_{t_{0}}^{t}\frac{dt'}{t'}
=\log \frac{t}{t_{0}}. 
\end{align}
The finite transformation of the original time coordinate is 
\begin{align}
\label{c1b2b}
t&=t_{0}e^{\tau}
\end{align}
and the dilatation $D$ rescales the time coordinate. 
It can be viewed as a global time dilation 
for the original time coordinate $t_{0}$ with a constant scale factor $e^{\tau}$. 
Its infinitesimal transformation is 
\begin{align}
\label{c1b2c}
\delta t&=t-t_{0}\approx \tau t_{0}.
\end{align}

\item local time dilation

The new time coordinate (\ref{c1a4}) 
whose evolution is generated by the special conformal transformation $K$ is 
\begin{align}
\label{c1b3a}
\tau&=\int_{t_{0}}^{t}\frac{dt'}{t^{'2}}
=\frac{1}{t_{0}}-\frac{1}{t}. 
\end{align}
The finite transformation of the original time coordinate is 
\begin{align}
\label{c1b3b}
t&=\frac{t_{0}}{1-\tau t_{0}}
\end{align}
and the generator $K$ 
is responsible for the local scale transformation. 
It corresponds to the local time dilation with the scale factor $\frac{1}{1-\tau t_{0}}$ depending 
on the time coordinate $t_{0}$.　
The infinitesimal transformation is
\begin{align}
\label{c1b3c}
\delta t&=t-t_{0}\approx \tau t_{0}^{2}. 
\end{align}
\end{enumerate}
A sequence of three finite transformations 
(\ref{c1b1b}), (\ref{c1b2b}) and (\ref{c1b3b}) can be expressed by 
\begin{align}
\label{cf1a1}
t&\rightarrow t'=f(t)=\frac{at+b}{ct+d}, &
A=\left(
\begin{array}{cc}
a&b\\
c&d\\
\end{array}
\right)\in SL(2,\mathbb{R}). 
\end{align}
The infinitesimal transformations 
(\ref{c1b1c}), (\ref{c1b2c}) and (\ref{c1b3c}) are summarized as
\begin{align}
\label{ci1a1}
\delta t&=\epsilon_{1}+\epsilon_{2}t+\epsilon_{3}t^{2}
\end{align}
where $\epsilon_{1}$, $\epsilon_{2}$ and $\epsilon_{3}$ 
are the infinitesimal parameters 
of the Hamiltonian $H$, 
the dilatation $D$ 
and the special conformal transformation $K$ 
respectively. 
The conformal generators obey the commutation relations
\begin{align}
\label{cc1a1}
[H,D]&=iH,& [K,D]&=-iK,& [H,K]&=2iD,
\end{align}
which form the $\mathfrak{sl}(2,\mathbb{R})$ algebra. 
In terms of the conformal generators, 
the Casimir operator $\mathcal{C}_{2}$ 
of the $\mathfrak{sl}(2,\mathbb{R})$ conformal algebra is written as 
\begin{align}
\label{cc1a2}
\mathcal{C}_{2}&=\frac12 (HK+KH)-D^{2}=KH+iD-D^{2}.
\end{align}
We claim that 
this expression specifies a choice of basis and its dual of the conformal algebra 
in terms of the Hamiltonian, the dilatation and the special conformal transformation of time coordinate $t$ 
although one can obtain an alternative quantum mechanical description with different time coordinate 
$t'=f(t)$ from (\ref{cf1a1}) and its Hamiltonian $H'$. 
For instance, 
for the DFF-model \cite{deAlfaro:1976je} with the action (\ref{hso1a}), 
one can find the theory with different Lagrangian 
\begin{align}
\label{dffnew1}
\mathcal{L}'&=\frac12 \left(
\dot{x}^{2}-\frac{g}{x^{2}}-\frac{x^{2}}{4}
\right)
\end{align}
containing the harmonic potential by changing the original time coordinate $t$ into a new time coordinate
\begin{align}
\label{dffnew2}
t'&=2\tan^{-1}t
\end{align}
whose Hamiltonian is 
\begin{align}
\label{dffnew3}
H'&=\frac12(H+K). 
\end{align}
This achieves the discrete energy spectrum and the normalizable ground state. 
However, this is different theory from the original one 
and the generators $H$, $D$ and $K$ should still be identified as the conformal generators in time coordinate $t$. 
Therefore the relation (\ref{cc1a2}) would intrinsically characterize 
the conjugation and scalar product in the state space of conformal quantum mechanics in time coordinate $t$.

One might worry that 
the system with time coordinate $t$ has neither discrete energy spectrum nor normalizable vacuum state, 
as observed in \cite{deAlfaro:1976je}. 
However, it would not imply that 
the physical system with time coordinate $t$ is unreasonable but rather that 
the quantization needs a subtle treatment due to the existing constraints on the canonical variables \cite{Strocchi:2016kce}. 
In other words, such undesirable properties for the physical description 
originate from a naive assumption in the quantization problem that 
all the canonical variables are the observables in the Hilbert space. 
In fact, for the DFF-model, the constraint $x>0$ should be taken seriously 
to proceed the consistent quantization so that 
some operators and states would not belong to the algebra of the observables 
and the physical states respectively 
\footnote{
The uncertain principal leads to the non-zero lowest energy for the physical ground states as the zero point energy 
although in the quantum field theories it is neglected by the normal ordering. 
In this aspect the non-normalizable vacuum state may play a role the reference state 
rather than the physical ground state. 
}. 
As a consequence, the non-normalizable vacuum state 
may be harmless so that one obtains descriptions of some physical systems 
\footnote{
For example, 
the energy spectrum of an electron in a periodic potential can be purely continuous 
and the ground state is not normalizable. 
However, by restricting the wavefunction to a unit cell, 
one can acquire the physically well-defined description. }.

In this work, 
instead of considering the detailed quantization prescription that 
provides a way to embed the observable algebra into a larger algebra of the canonical variables in a specific model, 
we wish to extract universal features of conformal quantum mechanics 
by restricting our attention just to the universal relation (\ref{cc1a2}).

\section{Vacuum State}
\label{vsubsec}
Let us consider the vacuum state $|\Omega\rangle$ 
which has zero energy
\begin{align}
\label{cqmvm1a}
H|\Omega\rangle&=0, & \langle \Omega|\Omega\rangle&=1. 
\end{align}
Here ket $|\cdot \rangle$ is the vector which belongs to a vector space $V$. 
The corresponding bra $\langle\cdot |$ is the linear map from $V$ to $\mathbb{C}$ 
which belongs to the space $V^{*}$ dual to $V$. 
Here we have assumed that 
the restrictions or boundary conditions of the wavefunctions in a particular region  
due to the constraints on the canonical variables admit the satisfactory normalizable wavefunctions 
for the vacuum states. 
Now that the states in quantum mechanics follow 
the $SL(2,\mathbb{R})$ conformal symmetry of time, 
one needs to select out consistent bra $\langle \cdot |$ belonging to the dual space $V^{*}$ 
in such a way that the matrix element 
$\langle \cdot |\mathcal{C}_{2}|\cdot \rangle$ 
of the Casimir operator (\ref{cc1a2}) 
gives the $c$-number. 
This leads to additional constraints on the bra-ket. 
From the expression (\ref{cc1a2}) of the Casimir operator 
the ket $H|\cdot\rangle$ has the dual bra $\langle\cdot |K$ 
so that $\langle \cdot|KH|\cdot\rangle$ gives a $c$-number. 
For the vacuum state (\ref{cqmvm1a}) 
one can choose the corresponding bra as 
\begin{align}
\label{cqmvm1b}
\langle \Omega|K&=0.
\end{align}
It follows from the $\mathfrak{sl}(2,\mathbb{R})$ algebra (\ref{cc1a1}) 
that the state 
\begin{align}
\label{cqmvm1d}
|\Omega'\rangle&:=D|\Omega\rangle
\end{align}
is also the vacuum state 
$H|\Omega'\rangle=0$.
According to the Casimir (\ref{cc1a2}), 
the ket $|\Omega'\rangle$ entails the bra 
\begin{align}
\label{cqmvm1d1}
\langle \Omega'|&\propto i\langle \Omega|,
\end{align}
which is proportional to $\langle \Omega|$. 
The bra-ket pairs from (\ref{cqmvm1d}) and (\ref{cqmvm1d1}) mean that 
an application of $D$ to $|\Omega\rangle$ does not result in 
a state vector with a distinct basis 
but just give a proportionality constant. 
Namely the vacuum state is the eigenstate of the dilatation $D$
\begin{align}
\label{cqmvm3a}
D|\Omega\rangle&=id|\Omega\rangle
\end{align}
where $d$ is the eigenvalue 
characterizing the scaling dimension of the vacuum $|\Omega\rangle$.  
In this work we will concentrate on 
the case with a real scaling dimension $d\in \mathbb{R}$ 
\footnote{
In contrast, the complex scaling dimension $d\in \mathbb{C}$ is realized for 
the principal series representation of 
the $\mathfrak{sl}(2,\mathbb{R})$ conformal algebra, 
in which case the energy eigenstate is described by the Whittaker vector 
\cite{Okazaki:2015lpa}. }. 
From (\ref{cc1a2}) the dual bra of the ket $D|\Omega\rangle$ 
is proportional to both $i\langle \Omega|$ and $\langle \Omega|D$. 
The proportionality constant is fixed by 
taking the bra-ket pairs from (\ref{cqmvm1d1}) and (\ref{cqmvm3a}). 
So we have 
\begin{align}
\label{cqmvm3b}
\langle \Omega|D&=id\langle \Omega|. 
\end{align}
Note that 
the dilatation generator is anti-hermitian 
and it is not the observable that is measured by a probability distribution. 
The diagonalization (\ref{cqmvm3a}) means that 
$|\Omega'\rangle$ is proportional to $|\Omega\rangle$ 
and there is a single basis $|\Omega\rangle$ of the vacuum state. 
From (\ref{cqmvm3a}) and (\ref{cqmvm3b}) one finds inner products
\begin{align}
\label{cqmvm3d}
\langle \Omega|D|\Omega\rangle&=id,\\
\label{cqmvm3f}
\langle \Omega|D^{2}|\Omega\rangle&=-(d+\mathcal{C}_{2})=-d^{2}.
\end{align}
It follows from (\ref{cqmvm3f}) that 
the scaling dimension $d$ of the vacuum is fixed by 
the Casimir $\mathcal{C}_{2}$
\begin{align}
\label{cqmv3g}
d&=\frac{1\pm\sqrt{1+4\mathcal{C}_{2}}}{2}.
\end{align}
Alternatively, the Casimir is expressed as 
\begin{align}
\label{cqmv3h}
\mathcal{C}_{2}&=d(d-1). 
\end{align}
Let us act $K$ on the vacuum and define
\begin{align}
\label{cqmvm2a}
|\Omega''\rangle&:=K|\Omega\rangle. 
\end{align}
From the $\mathfrak{sl}(2,\mathbb{R})$ algebra (\ref{cc1a1}) 
and the diagonalization (\ref{cqmvm3a}) we see that it satisfies the relations
\begin{align}
\label{cqmv4a}
H|\Omega''\rangle&=
-2d|\Omega\rangle,\\
\label{cqmv4b}
D|\Omega''\rangle&=
i(d+1)|\Omega''\rangle. 
\end{align}
(\ref{cqmv4b}) means that 
$K$ increases the eigenvalue of $|\Omega\rangle$ for $D$ 
by $i$. 
For a further application of $K$ on the vacuum 
we find 
\begin{align}
\label{cqmv4c}
D\left(
K^{2}|\Omega\rangle
\right)
&=
i(d+2)\left(
K^{2}|\Omega\rangle. 
\right) 
\end{align}
This implies that 
$K^{2}$ increases the eigenvalue of $|\Omega\rangle$ for $D$ by $2i$.

As a simple example, let us consider the DFF-model \cite{deAlfaro:1976je} 
whose action is given by
\begin{align}
\label{hso1a}
S&=\frac12 \int dt 
\left(\dot{x}^{2}-\frac{g}{x^{2}}\right)
\end{align}
where $g$ is a dimensionless coupling constant parameter. 
The action (\ref{hso1a}) is invariant under the conformal transformations 
(\ref{cf1a1}) and $\delta x=x/(ct+d)$. 
Using the Noether method, 
one can deduce the conformal generators 
\begin{align}
\label{dff1a1}
H&=\frac{p^{2}}{2}+\frac{g}{2x^{2}},& 
D&=-\frac14\{x,p\},& 
K&=\frac12 x^{2}
\end{align}
and the Casimir operator
\begin{align}
\label{hso1b}
\mathcal{C}_{2}&=d(d-1)
=\frac{g}{4}-\frac{3}{16}.
\end{align}
Therefore if the vacuum state exists in the DFF-model, 
the coupling constant $g$ determines the scaling dimension of the vacuum 
\begin{align}
\label{hso1c}
d&=\frac{1\pm\sqrt{g+\frac14}}{2}. 
\end{align} 
For example, the Heisenberg picture vacuum is realized when $g=\frac34$.

\section{$D$-function}
\label{dsec}
Consider an energy eigenstate $|E\rangle$ 
\begin{align}
\label{cqmen1a}
H|E\rangle&=E|E\rangle
\end{align}
with energy eigenvalue $E\in \mathbb{R}$. 
Taking into account the hermiticity of the Hamiltonian $H$ 
and the expression (\ref{cc1a2}) of the Casimir operator 
we can take the corresponding bra for the state (\ref{cqmen1a}) as
\begin{align}
\label{cqmen1a1}
\langle E|K&=\langle E|E
\end{align}
where $\langle E|$ is the dual bra of the energy eigenstate $|E\rangle$. 
Let us apply $D$ and $K$ to the energy eigenstate and define
\begin{align}
\label{cqmen1b}
|E'\rangle&:=D|E\rangle,&  
|E''\rangle&:=K|E\rangle. 
\end{align}
Then we have
\begin{align}
\label{cqmen1c}
H|E'\rangle&=E|E'\rangle+iE|E\rangle,\\
\label{cqmen1d}
H|E''\rangle&=E|E''\rangle+2i|E'\rangle.
\end{align}
From (\ref{cqmen1c}) 
the state $|E'\rangle$ is not the energy eigenstate 
due to the term $iE|E\rangle$. 
The energy eigenstate $|E\rangle$ is unchanged under the scale transformation 
generated by $D$ only when $|E\rangle$ is the vacuum state $|\Omega\rangle$ 
\footnote{
This fact was also pointed out in Appendix C of \cite{Pal:2016rpz}. 
}. 

Similarly according to (\ref{cqmen1d}), 
$|E''\rangle$ is not the energy eigenstate 
because of the term $2i|E'\rangle$. 
The energy eigenstate can be realized under the conformal boost 
generated by $K$ as the energy eigenstate 
only when $|E'\rangle$ vanishes, i.e. $d=0$ and $E=0$. 
In this case (\ref{cqmen1d}) requires that 
$|E''\rangle$ is the vacuum state. 
Then (\ref{cqmen1b}) implies that 
$|E''\rangle$ is the eigenstate of $K$ 
\begin{align}
\label{cqmen1d1}
K|\Omega\rangle&=k|\Omega\rangle 
\end{align}
with eigenvalue $k$. 
Therefore only the vacuum state obeying (\ref{cqmen1d1}) and 
\begin{align}
\label{cqmen1e}
H|\Omega\rangle&=D|\Omega\rangle=0
\end{align}
keeps the same energy eigenvalue under the conformal transformations. 
In particular the conformally invariant vacuum is realized 
only when the vacuum satisfies
\begin{align}
\label{cqmen1f}
H|\Omega\rangle&=D|\Omega\rangle=K|\Omega\rangle=0.
\end{align}
In this case the vacuum state admits the Heisenberg picture 
in which the state has no time dependence.

Employing the Baker-Campbell-Hausdorff formula, 
we find that 
\begin{align}
\label{app1a}
e^{H}De^{-H}&=D+iH,& e^{H}Ke^{-H}&=K+2iD-H,\\
\label{app1b}
e^{aD}He^{-aD}&=e^{-ia}H,& e^{aD}Ke^{-aD}&=e^{ia}K. 
\end{align}
Using the relation (\ref{app1b}), 
one can show that 
\begin{align}
\label{app1d}
H\left(
e^{i\alpha D}|E\rangle 
\right)&=e^{-\alpha}E
\left(
e^{i\alpha D}|E\rangle 
\right). 
\end{align}
Taking $\alpha$ as 
a continuous parameter, 
the energy spectrum can be continuous. 
Hence the continuous energy spectrum is a universal feature 
in conformal quantum mechanics. 
However, such undesirable feature can be cured by selecting the observables out of the canonical operators. 
Thus it does not conclude that 
one should discard the system with time coordinate $t$ as the physical system.

Now consider a matrix element
\begin{align}
\label{eco1b}
D(E):=\frac{1}{i}\langle E|D|E\rangle. 
\end{align}
This describes the quantum scaling dimension 
of the energy eigenstate $|E\rangle$ 
and it is a real function of the energy eigenvalue $E$. 
The overall factor in (\ref{eco1b}) eliminates the imaginary unit due to the anti-hermiticity of 
the dilatation generator $D$. 

We assume that 
the energy eigenstate $|E\rangle$ forms a complete orthonormal set 
\footnote{
In \cite{Jackiw:2012ur}, for the DFF-model 
the energy eigenstates $|E\rangle$ with these properties are actually constructed from the proposed state $|t\rangle$ 
by the Fourier transform 
\begin{align}
\label{tstate1a}
|E\rangle&=
2^{r_{0}}E^{\frac12-r_{0}}
\int_{-\infty}^{\infty}
\frac{dt}{2\pi}e^{-iEt}|t\rangle. 
\end{align}
The author thanks R. Jackiw for pointing out this point. 
}
\begin{align}
\label{e1a}
1&=\int dE |E\rangle \langle E|,& 
\langle E_{1}|E_{2}\rangle&=\delta(E_{1}-E_{2}). 
\end{align}
Making use of 
(\ref{cc1a2}), (\ref{cqmen1a}), (\ref{cqmen1a1}) and (\ref{e1a}), 
we find the quadratic equation
\begin{align}
\label{dfcn1a0}
D(E)^{2}-D(E)-(\mathcal{C}_{2}-E^{2})&=0, 
\end{align}
whose solution is given by 
\begin{align}
\label{dfcn1a}
D(E)&=
\frac{1\pm\sqrt{1+4(\mathcal{C}_{2}-E^{2})}}{2}. 
\end{align}

To make all predictions in quantum mechanics work correctly, 
we shall associate some energy eigenstate $|E\rangle$ of the energy $E$ with the unitary group 
to describe time evolution, i.e. the unitary evolution. 
However, as the quantity $D(E)$ measures the averaged scaling dimension of the energy eigenstate $|E\rangle$,  
the energy eigenstate $|E\rangle$ would behave as $t^{D(E)}$. 
So it is preferable to have $D(E)=0$. 
To achieve this, we will need to take the minus sign in (\ref{dfcn1a}) 
and we have
\begin{align}
\label{dfcn2a}
D(E)&=\frac{1-\sqrt{1+4(\mathcal{C}_{2}-E^{2})}}{2},
\end{align}
which we will call a $D$-function.

\begin{figure}
\begin{center}
\includegraphics[width=6.5cm, angle=90]{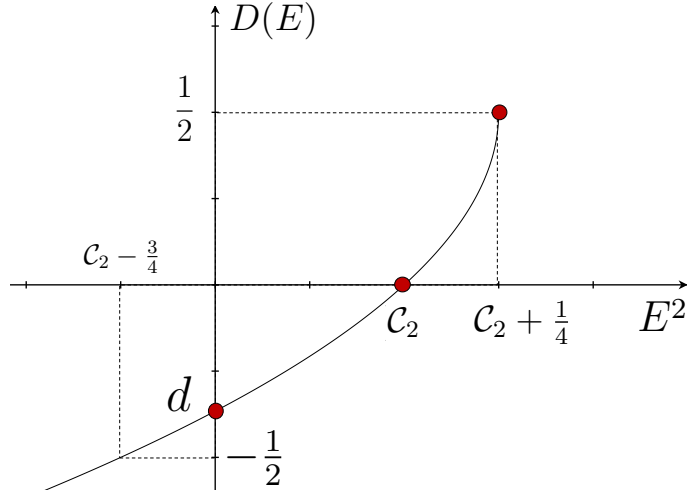}
\caption{The $D$-function. 
It is defined in the range $0\le E^{2}\le \mathcal{C}_{2}+\frac14$ and 
decreasing monotonically from the UV to the IR. }
\label{figdfn}
\end{center}
\end{figure}

Now let us make a connection to the AdS/CFT correspondence. 
It tells \cite{Gubser:1998bc,Witten:1998qj} that the bulk mass $m$ of a scalar field in AdS$_{2}$ space  
is related to the dimension $\Delta_{m}$ of the corresponding operator 
on the boundary as
\begin{align}
\label{ads2a1}
\Delta_{m}(\Delta_{m}-1)&=m^{2}
\end{align}
and there are two solutions 
\begin{align}
\label{ads2a2}
\Delta_{m}^{\pm}&=\frac{1\pm \sqrt{1+4m^{2}}}{2}.
\end{align}
For $\frac34<m^{2}$ 
only the boundary conditions with $\Delta_{m}^{+}$ lead to the normalizable solution 
\cite{Breitenlohner:1982bm, Breitenlohner:1982jf, Mezincescu:1984ev}
as $z^{\Delta_{m}^{+}}$ near $z=0$ for a free scalar of mass $m$ 
in the AdS$_{2}$ space whose metric is given by
\begin{align}
\label{ads2}
ds^{2}&=\frac{1}{z^{2}}(dz^{2}+dt^{2}). 
\end{align}
Since (\ref{dfcn2a}) corresponds to $\Delta_{m}^{-}$, 
it is unlikely that the dual conformal quantum mechanics appears when $\frac34<m^{2}$. 
Meanwhile there can be two possible boundary conditions 
with $\Delta_{m}^{+}$ and $\Delta_{m}^{-}$ 
when \cite{Breitenlohner:1982bm, Breitenlohner:1982jf, Mezincescu:1984ev} 
\begin{align}
\label{bfwindow1}
-\frac14<m^{2}<\frac34
\end{align}
where the lower bound is the Breitenlohner-Freedman bound 
\footnote
{It has been discussed \cite{Pioline:2005pf} that 
an electric field $\mathcal{E}$ in AdS$_{2}$ can shift 
the Breitenlohner-Freedman bound 
$-\frac14<m^{2}$ to $-\frac14+\mathcal{E}^{2}\le m^{2}$ 
due to the pair production of the Schwinger effect. 
But we will not consider such effect in this work.}. 
Comparing (\ref{dfcn1a}) with (\ref{ads2a2}) for $\Delta_{m}^{-}$, 
the mass range leads to 
$E^{2}-\frac14\le \mathcal{C}_{2}\le E^{2}+\frac34$. 
The existence of the vacuum state requires that the Casimir 
is bounded above and below; 
$-\frac14\le \mathcal{C}_{2}\le \frac34$ 
and that the scaling dimension $d$ of the vacuum has a bound 
$-\frac12\le d\le \frac12$. 
The resulting function (\ref{dfcn2a}) is shown in Figure \ref{figdfn}. 
As the energy eigenvalue $E$ takes real values, 
the energy squared is bounded above and below; 
$0\le E^{2}\le \mathcal{C}_{2}+\frac14$. 
Correspondingly, $D(E)$ is also bounded above and below
\begin{align}
\label{dfcn2b}
d\le D(E)\le \frac12
\end{align}
where $d$ is the scaling dimension (\ref{cqmv3g}) of the vacuum state. 
The normalizability of the energy eigenstate under the evolution operator can be kept 
when $E^{2}=\mathcal{C}_{2}$. 

To summarize, 
if conformal quantum mechanics dual to the AdS$_{2}$ 
has the vacuum $|\Omega\rangle$ and a complete orthonormal set of energy eigenstates $|E\rangle$ 
with the averaged scaling dimension (\ref{dfcn2a}), 
the energy eigenstate will realize well-behaved unitary evolution at $E^{2}=\mathcal{C}_{2}$ 
for $-\frac14<\mathcal{C}_{2}<\frac34$.

Here we would like to argue the physical implications of the $D$-function. 
It involves two important physical quantities, scaling dimension and energy. 
In quantum field theory change of scale is described by the RG transformation 
and the number of degrees of freedom in a physical system decreases along the RG flows 
from high energy to lower energy.  
In CFT$_{2}$ this can be quantitatively measured by defining a $c$-function \cite{Zamolodchikov:1986gt} 
which has the following properties:  
\begin{enumerate}
\item It is a real function of coupling constant $g$ and energy scale $E$ which is defined on the space of theories. 
\item It monotonically decreases along the RG flow. 
\item It is stationary at the RG fixed point where its value equals the crucial parameter in the CFT. 
\end{enumerate}
It is proposed \cite{Cardy:1988cwa} that 
for even dimensional CFT$_{d}$, 
the Euler characteristic appearing 
in the trace anomaly provides a $c$-function,  
which can be evaluated as the expectation value of the trace of the energy-momentum tensor on the sphere $S^{d}$
\begin{align}
\label{cfn1}
c\sim \int_{S^{d}}\langle T_{\mu}^{\mu}\rangle. 
\end{align}
We remark that the trace $T^{\mu}_{\mu}$ of the energy momentum tensor is related to 
the conserved current of the dilatation $D$  
as the scale invariance is achieved  
when the trace of the energy momentum tensor vanishes 
\cite{Coleman:1970je}. 
This leads us to expect that 
in conformal quantum mechanics 
the expectation value $\langle D\rangle$ of the dilatation, 
which depends on the energy scale $E$, 
can play a role of a $c$-function. 
In fact, we see that the $D$-function has the above properties of a $c$-function: 
\begin{enumerate}
\item 
The $D$-function is defined on the space of theory 
as it depends on the Casimir which may involve the coupling constants of the theories considered. 
For instance, in the DFF-model (\ref{hso1a}) the Casimir invariant is given by the 
coupling constant $g$ as in (\ref{hso1c}) and the $D$-function is represented by 
\begin{align}
\label{dfn3a}
D(E)&=\frac{1-\sqrt{g-4E^{2}+\frac14}}{2}. 
\end{align}
Since the dimensionless coupling constant $g$ parameterizes the theories, 
the $D$-function is defined on the space of the theories. 

\item Along the flow from the UV to the IR, 
the energy scale decreases 
and the $D$-function decreases monotonically with the energy scale $E$
\begin{align}
\label{dfn3b}
\frac{dD(E)}{dE}
&=\frac{2E}{\sqrt{1+4(\mathcal{C}_{2}-E^{2})}}\ge 0.
\end{align}
This shows the monotonic flow for the $D$-function. 

\item At the fixed point $E=0$ of the flow, 
it is stationary with its value 
\begin{align}
\label{dfcn1b}
D(E=0)&=d=\frac{1-\sqrt{1+4\mathcal{C}_{2}}}{2}. 
\end{align}
This is the crucial parameter in conformal quantum mechanics, 
that is the scaling dimension (\ref{hso1a}) of the vacuum state. 
\end{enumerate}
Therefore the $D$-function exhibits analogous properties as a $c$-function.  
It supports the irreversibility of the flow from the UV to the IR 
in conformal quantum mechanics. 
At the fixed point it becomes the scaling dimension $d$ of the vacuum 
that encodes the theory considered.  
As dimension conceptually measures certain properties of an object 
that is independent from other objects, 
the $D$-function as an averaged scaling dimension of energy eigenstates, 
including the scaling dimension $d$ of the vacuum counts 
the number of degrees of freedom in a similar way as a $c$-function.

\section{Bounds on Scaling Dimensions}
\label{nogosec}
Now we want to consider a dynamical realization 
of conformal group in quantum mechanics. 
We shall postulate the existence of primary operators 
transforming as representations of the conformal algebra
\begin{align}
\label{op1a1}
\left(
T(g)\mathcal{O}
\right)_{\alpha}(t)&=S_{\alpha\beta}(g,t)\mathcal{O}_{\beta}(g^{-1}t)
\end{align}
where $g$ acts on time coordinate $t$ as (\ref{cf1a1}) 
and $T(g)$ is the representation matrix. 
It follows from (\ref{op1a1}) that 
$S_{\alpha\beta}(g,0)$ should be a representation of 
the stability subgroup at time $t=0$. 
According to the infinitesimal transformation (\ref{ci1a1}) 
this subgroup is given by the dilatation and special conformal
transformation. 
The commutation relation (\ref{cc1a1}) reduces to 
\begin{align}
\label{op1a2}
[K,D]&=-iK
.
\end{align}
Every element of the $\mathfrak{sl}(2,\mathbb{R})$ 
conformal algebra can be constructed 
by ascribing the time dependence to the generators. 
From (\ref{app1a}) we have
\begin{align}
\label{op1a3}
D(t)&=e^{iHt}De^{-iHt}=D-tH,\\
\label{op1a4}
K(t)&=e^{iHt}Ke^{-iHt}=K-2tD+t^{2}H
\end{align}
Assume that 
\begin{align}
\label{op1b1}
H\mathcal{O}(0)&=i\dot{\mathcal{O}}(0),\\
\label{op1b2}
D\mathcal{O}(0)&=i\Delta\mathcal{O}(0),\\
\label{op1b3}
K\mathcal{O}(0)&=0
\end{align}
with $\Delta\in\mathbb{R}$. 
(\ref{op1b2}) says that the operator $\mathcal{O}(0)$ enjoys a real scaling dimension $\Delta$. 
According to (\ref{op1a3}) and (\ref{op1a4}) 
we find that 
\begin{align}
\label{op1b5a}
H\mathcal{O}_{\Delta}(t)&=i\dot{\mathcal{O}}_{\Delta}(t),\\
\label{op1b5b}
D\mathcal{O}_{\Delta}(t)&=
i\left(-t\frac{\partial}{\partial t}+\Delta
\right)\mathcal{O}_{\Delta}(t),\\
\label{op1b5c}
K\mathcal{O}_{\Delta}(t)&=
i\left(
t^{2}\frac{\partial}{\partial t}-2t\Delta
\right)\mathcal{O}_{\Delta}(t). 
\end{align}
Equivalently one can define the primary operators $\mathcal{O}_{\Delta}(t)$ 
which obey (\ref{op1b5a})-(\ref{op1b5c}) by the transformation law
\begin{align}
\label{op1b6a}
\mathcal{O}_{\Delta}(t)&\rightarrow 
\left(
\frac{\partial t'}{\partial t}
\right)^{\Delta}\mathcal{O}_{\Delta}(t')
=\frac{1}{(ct+d)^{2\Delta}}\mathcal{O}_{\Delta}(t')
\end{align}
under the finite transformation (\ref{cf1a1}).

We will formulate conformal quantum mechanics 
in terms of the primary operators $\mathcal{O}_{\Delta}(t)$ 
acting on the vacuum state $|\Omega\rangle$. 
We assume that each state in the Hilbert space 
is represented by 
\begin{align}
\label{state1a}
|\mathrm{state}\rangle 
&=
F(G)|\mathcal{O}_{\Delta_{1}}(t_{1})\cdots\mathcal{O}_{\Delta_{n}}(t_{n})\rangle
\end{align}
where
\begin{align}
\label{state1b}
|\mathcal{O}_{\Delta_{1}}(t_{1})\cdots\mathcal{O}_{\Delta_{n}}(t_{n})\rangle
&=\mathcal{O}_{\Delta_{1}}(t_{1})\cdots\mathcal{O}_{\Delta_{n}}(t_{n})|\Omega\rangle
\end{align}
with $F(G)$ being some function of $G=uH+vD+wK$. 
Let us examine the expectation values 
$\langle \mathrm{state} A|\mathrm{state} B\rangle$ 
constructed as overlaps 
of the two states $|\mathrm{state} A\rangle$ 
and $|\mathrm{state} B\rangle$ 
with the form of (\ref{state1a}) in the Hilbert space. 
In this work we will explore the expectation value involving the time-independent primary operators 
$\mathcal{O}_{\Delta}:=\mathcal{O}_{\Delta}(0)$ 
and take the conventional choice of the overall constant one which 
fixes the normalization of $\mathcal{O}_{\Delta}$ as
\begin{align}
\label{state2a}
\langle \mathcal{O}_{\Delta}|\mathcal{O}_{\Delta}\rangle &=1. 
\end{align}

\begin{figure}
\begin{center}
\includegraphics[width=8cm]{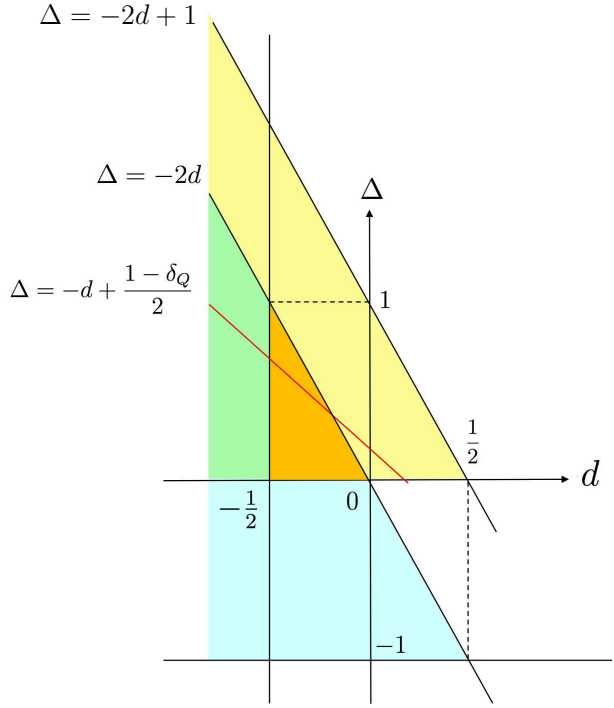}
\vspace*{1.5cm}
\caption{The bound on $(d,\Delta)$ in conformal quantum mechanics. 
In the orange region the primary operators with the dimension $\Delta$ 
and the vacuum states with the dimension $d$ are allowed 
and the red line characterizes the charged physical states coupled to the gauge operators of $\delta_{Q}$. 
Without the favored condition (\ref{nogo1a2}) for the unitary evolution of the states, 
the green region is also allowed. }
\label{fignogo}
\end{center}
\end{figure}

Now we would like to extract constraints on the description 
of the unitary evolution for a certain physical system. 
To achieve this, one needs to fix its time coordinate $t$ and construct all the physical states 
in such a way that they fall into the representations of the $\mathfrak{sl}(2,\mathbb{R})$ conformal algebra 
specified by the vacuum with the eigenvalue of the Casimir invariant $\mathcal{C}_{2}$, 
i.e. the scaling dimension $d$. 
Given the normalized primary operators (\ref{state2a}), 
this corresponds to the condition
\begin{align}
\label{en1}
\langle \mathcal{O}_{\Delta}|\mathcal{C}_{2}|\mathcal{O}_{\Delta}\rangle 
&=\mathcal{C}_{2}=d(d-1), 
\end{align}
which ensures the unitary evolution of the states by fixing the eigenvalue of the Casimir invariant. 
Alternatively, we can write the expectation value (\ref{en1}) as 
\begin{align}
\label{en2}
\langle \mathcal{O}_{\Delta}|HK-iD-D^{2}|\mathcal{O}_{\Delta}\rangle 
&=\langle \mathcal{O}_{\Delta}|HK|\mathcal{O}_{\Delta}\rangle 
+(d+\Delta)(d+\Delta+1). 
\end{align}
Unitarity implies the positivity of the inner product in the Hilbert space. 
Demanding that 
$\langle \mathcal{O}_{\Delta}|HK|\mathcal{O}_{\Delta}\rangle=
|K|\mathcal{O}_{\Delta}\rangle |^{2}$ is positive definite 
and combining (\ref{en1}) with (\ref{en2}), 
we find a condition 
\begin{align}
\label{en3}
(\Delta+2d)(\Delta+1)\le 0. 
\end{align}
Together with the preferred range (\ref{dfcn2b}) 
under the unitary evolution probed by the $D$-function, 
we obtain the bounds on scaling dimension 
of the primary operator and of the vacuum 
\begin{align}
\label{nogo1a1}
-1&\le \Delta \le -2d, \\
\label{nogo1a2}
-\frac12&\le d\le \frac12.
\end{align}
Similarly we can extract further constraints 
by rewriting (\ref{en1}) as 
\begin{align}
\label{en4}
\langle \mathcal{O}_{\Delta}|KH+iD-D^{2}|\mathcal{O}_{\Delta}\rangle
&=\langle \mathcal{O}_{\Delta}|KH|\mathcal{O}_{\Delta}\rangle
+(d+\Delta)(d+\Delta-1).
\end{align}
Since $\langle
\mathcal{O}_{\Delta}|KH|\mathcal{O}_{\Delta}\rangle
=|H|\mathcal{O}_{\Delta}\rangle|^{2}$ is positive definite, 
we get a condition 
\begin{align}
\label{en5}
\Delta(\Delta+2d-1)&\le 0, 
\end{align}
which gives the additional constraint
\begin{align}
\label{nogo1a1a}
0\le \Delta \le -2d+1. 
\end{align}
The result is depicted in Figure \ref{fignogo}. 
The primary operators and the vacuum states can exist 
in the orange region. 
As a consequence, the allowed range of the scaling dimensions of the 
physical states $|\mathcal{O}_{\Delta} \rangle$ which are constructed in terms of 
the vacua and the primary operators is
\begin{align}
\label{nogo1a1}
-\frac12\le d+\Delta\le \frac12. 
\end{align}
It supports the existence of the bosonic scalar with scaling dimension $-\frac12$, 
the fermion with the scaling dimension $0$ and the bosonic auxiliary field with scaling dimension $\frac12$ 
in conformal quantum mechanics, 
as argued and constructed in the Lagrangian theory. 
If we relax the condition (\ref{nogo1a2}) for the favored energy eigenstates $|E\rangle$ under the unitary evolution, 
which is examined by the $D$-function, the states are allowed in the green region.

Suppose that 
a theory allows the construction of a conserved charge. 
In what follows, we will not rely on the Lagrangian, but rather 
describe a charge as the operator that acts on the state (\ref{state1a}) and the primary operator (\ref{op1b6a}).   
Let $Q$ be the corresponding charge operator that obeys 
\begin{align}
\label{nogo2a1}
Q\mathcal{O}_{\Delta}&=q\mathcal{O}_{\Delta},\\
\label{nogo2a2}
Q|\Omega\rangle&=0,\\
\label{nogo2a3}
[H,Q]&=0,\\
\label{nogo2a4}
[D,Q]&=i\delta_{Q}Q
\end{align}
with $q\in \mathbb{R}$. 
(\ref{nogo2a1}) and (\ref{nogo2a2}) assign the charges 
such that the primary operator $\mathcal{O}_{\Delta}$ has charge $q$ whereas the vacuum state has no charge. 
(\ref{nogo2a3}) implies that 
the charge operator $Q$ is not dynamical. 
In the Lagrangian description it would have no kinetic term, 
so it can be eliminated by its algebraic equation of motion as an auxiliary field. 
(\ref{nogo2a4}) gives the scaling dimension $\delta_{Q}$ of the charge operator $Q$. 
In the following we will focus on the case $\delta_{Q}\ge0$.

The corresponding symmetry transformation is a global transformation 
if $\delta_{Q}=0$ because every charge 
at $t$ is transformed in the same way so that 
$q$ is a constant charge, which we will call a global charge. 
For the continuous symmetry an operator can be realized 
by exponentiating the corresponding global charge. 
When the theory is generalized 
by including a $d-1$-dimensional space $M_{d-1}$ separated from time, 
one may further define higher-form global symmetries and higher-form global charges 
$Q(M_{d-1})$ \cite{Gaiotto:2014kfa} 
by furnishing the scaling dimensions stemming from $M_{d-1}$.

On the other hand, for $\delta_{Q}> 0$ 
we view the symmetry transformation as a local transformation 
because the charge is a function of time; $q(t)$. 
In this case $Q$ can enter the Lagrangian 
and its elimination by the equation of motion and the gauge fixing 
would gives the Gauss constraint. 
In the Lagrangian description of quantum mechanics, 
it is nothing but an auxiliary gauge field. 
In quantum mechanics, it is the Gauss law operator. 
Note that the Gauss law constraint is not the identity between 
operators obeying the canonical commutation relation $[\cdot, \cdot]$ 
but rather it holds only when acting on the physical states. 
In fact, it is well known that 
the Gauss law constraint is incompatible with the canonical commutation relation 
\footnote{
For example, in pure Maxwell theory the canonical commutation relations for the gauge fields 
\begin{align}
\label{qed1}
[A_{i}(\bm{x},0),\dot{A}_{j}(\bm{x}',0)]&=i\delta_{ij}\delta(\bm{x}-\bm{x}'),& 
[A_{i}(\bm{x},0),A_{j}(\bm{x}',0)]&=0
\end{align}
are incompatible with the Gauss law constraints 
\begin{align}
\label{qed2}
\mathrm{div} E&=0,& E_{i}&=\dot{A}_{i}. 
\end{align}
See for example \cite{Strocchi:2016kce} for a more general discussion on the Gauss law constraint in quantum mechanics. 
}. 
Therefore we should not require the Jacobi identity 
for a canonical commutation relation operation by including the Gauss law operator $Q$. 
When the theory is generalized 
by adding a $d-1$-dimensional space $M_{d-1}$, 
this operator behaves as a vector- or tensor-like operator 
since it has the non-vanishing scaling dimension. 
If a theory follows the action principle, 
it naturally appears in the covariant derivative as a gauge field 
to make the symmetry manifest. 
We will refer to the charge operator with $\delta_{Q}>0$ as a gauge operator.

Consider a matrix element
\begin{align}
\label{nogo2b1}
\langle \mathcal{O}_{\Delta}|[K,Q]H|\mathcal{O}_{\Delta}\rangle. 
\end{align}
Since $[K,Q]H=(KH)Q-Q(KH)$ and 
the both actions of $Q$ on the ket $|\mathcal{O}_{\Delta}\rangle$ 
and on the bra $\langle \mathcal{O}_{\Delta}|$ produce the same charge $q\in \mathbb{R}$, 
this should vanish. 
On the other hand, using the commutation relations 
(\ref{cc1a1}) and (\ref{nogo2a1})-(\ref{nogo2a4}), 
we find 
\begin{align}
\label{nogo2b2a}
[K,Q]H&=2i\delta_{Q}QD-\delta_{Q}^{2}Q+\delta_{Q}Q. 
\end{align}
Plugging this into (\ref{nogo2b1}) we get 
\begin{align}
\label{nogo2b2}
q\delta_{Q}\left(
\delta_{Q}-1+2(d+\Delta)
\right)&=0.
\end{align}
For the global charge operator with $\delta_{Q}=0$ 
the above condition holds and there is no constraint on the primary operator. 
However, for the gauge operator with $\delta_{Q}>0$ 
the scaling dimension of the charged primary operator is determined by 
\begin{align}
\label{nogo2c3}
d+\Delta&=\frac{1-\delta_{Q}}{2}. 
\end{align} 
The resulting constrained scaling dimension is illustrated in 
Figure \ref{fignogo}. 
The red line characterizes the gauge operator. 
Within the regions (\ref{nogo1a1}) and (\ref{nogo1a2}), 
there exists a bound on the scaling dimension of the gauge operator 
\begin{align}
\label{nogo2c4}
0< \delta_{Q}\le 2. 
\end{align}
This admits the presence of 
the gauge operators with $\delta_{Q}=1$, 
which would realize massless spin $s=1$ gauge fields involving photon and gluon, 
coupled to the physical states with $d+\Delta =0$, i.e. free fermions. 
Also it is compatible with the gauge operators with $\delta_{Q}=2$, 
which would show up as massless spin $s=2$ fields involving graviton, 
coupled to the physical states with $d+\Delta=-\frac12$, i.e. free bosonic scalars. 
On the other hand, the bosonic auxiliary field with $d+\Delta=\frac12$ 
may not couple to the gauge operators.

\section{Discussion}
\label{dissec}
In this work we have studied conformal quantum mechanics 
with the vacuum state and the primary operators. 
We have shown that 
a matrix element of the dilatation operator 
between two energy eigenstates may define a 
conformal quantum mechanical counterpart of a $c$-function, 
which we call a $D$-function. 
Its monotonic decrease from the UV to the IR 
along the flow supports the universal irreversibility 
of the RG flow in higher dimensional field theories. 
At the fixed point of the flow 
it becomes a crucial parameter $d$, that is the scaling dimension of the vacuum, 
which specifies the theory, analogous to the central charge 
in two-dimensional conformal field theories. 
In addition, we have found new no-go theorems 
which impose constraints and bounds on scaling dimensions 
of the primary operator, the vacuum and the gauge operators.


Our results in conformal quantum mechanics should have implications 
for two-dimensional gravity and black holes via holography. 
It would be nice to find further applications of our proposed $D$-function 
in a holographic framework of the RG flow 
as in higher dimensional conformal field theories in the context of the AdS/CFT correspondence 
\cite{Freedman:1999gp,Myers:2010xs}. 
Also our result may be substantiated by the dS/CFT correspondence \cite{Strominger:2001pn}. 
As the radial direction corresponds to time evolution in an asymptotically de Sitter space-time \cite{Strominger:2001pn, Balasubramanian:2002zh}, 
time evolution would be dual to the RG flow. 
It is discussed \cite{Halyo:2001cu} that 
the RG flow may correspond to the instability of the space due to the Hawking emission \cite{Hawking:1974sw}. 
When time goes on, the black hole mass would decrease due to the Hawking radiation. 
Nevertheless, Bianchi and Serlak \cite{Bianchi:2014qua,Bianchi:2014vea} 
have recently derived a formula that relates 
the energy flux and the von Neumann entropy of the Hawking radiation 
for a two-dimensional black hole, which predicts the possibility of the increase of black hole mass due to the negative energy flux, 
whereas Abdolrahimi and Page \cite{Abdolrahimi:2015tha} 
have pointed out some inadequacies in the Bianchi-Smerlak formula. 
We would like to address these issues via the holographic method in the future work. 

Although we have investigated several properties of conformal quantum mechanics, 
there would be other conformal quantum mechanical models which are beyond the scope of this work.  
Firstly one can construct other conformal quantum mechanics by considering topological quantum mechanics, 
which may not obey the unitary evolution for energy eigenstates in our argument. 
Such theories may have zero Hamiltonian, however, they can appears 
in the study of topological or BPS protected sector for physical system. 
For example superconductor and fractional Quantum Hall systems are studied by 
topological Chern-Simons quantum mechanics \cite{Manton:1997tg, Polychronakos:2001mi, Dorey:2016mxm} 
and a certain BPS protected sectors are examined by superconformal quantum mechanics \cite{Fubini:1984hf} 
(see also \cite{BrittoPacumio:1999ax, Fedoruk:2011aa, Okazaki:2015pfa} and references therein). 
By relaxing the unitary evolution of physical energy eigenstates, 
which is taken as the physically preferable conditions in this work, one may find other conformal quantum mechanics which can be used as a physical description. 

Meanwhile, even if we focus on the $SL(2,\mathbb{R})$ conformal quantum mechanics, there may exist some  quantum mechanical models which are not studied in this work. 
One possibility is to construct conformal quantum mechanics which does not rely on the existence of vacuum state and of the primary operators. 
In fact, Chamon, Jackiw, Pi and Santos \cite{Chamon:2011xk, Jackiw:2012ur} argue that 
in the DFF model the unexpected non-primary operator and the non-conformally invariant state conspire with each other 
to construct the consistent correlation functions.  

For many applications of the conformal quantum mechanics, it would be important to proceed with the analysis of correlation functions 
and to explore stronger constraints by considering additional physical requirement 
or the presence of additional operators 
characterizing symmetries in the theory.

\subsection*{Acknowledgements}
The author would like to thank 
C. Cardona, R. Jackiw, G. Dimitrios, D. S. Gupta, 
K. Hosomichi, S. Kawamoto, Y. Matsuo, S. Pal, D. J. Smith 
and S. Yamaguchi for useful discussions and comments. 
The author also thanks the organizers of the 
\textit{Workshop on String and M-theory in Okinawa} 
in Okinawa for providing a stimulating environment 
during the completion of the work. 
The author is supported by MOST under the Grant No.105-2811-066. 

\bibliographystyle{utphys}
\bibliography{ref}

\end{document}